\documentclass[useAMS,usenatbib]{mn2e}
\usepackage{epsfig}
\usepackage{subfigure}

\title[``Fundamental-Plane''-Like Relations]
     {``Fundamental-Plane''-Like Relations From
         Collisionless Stellar Dynamics:
         A Comparison of Mergers and Collapses}

\author[C. C. Dantas et al.]
{C.C. Dantas$^{1,2}$, H.V. Capelato$^{2}$, A.L.B. Ribeiro$^{3,4}$, and R.R.
de Carvalho$^{5,6}$\\
$^1$ Departamento de Astronomia,
Instituto Astron\^omico e Geof\'{\i}sico,
Universidade de S\~ao Paulo, 01060-970, SP, Brazil\\
$^2$ Divis\~ao de Astrof\'{\i}sica, INPE/MCT
CP 515, 12201-970 SP, Brazil\\
$^3$ Departamento de Matem\'atica Aplicada,
Universidade Estadual de Campinas, 13083-970 SP, Brazil\\
$^4$ {\it Present address:} DCET, Universidade Estadual de Santa Cruz, 
45650-000, Ilheus-BA, Brazil. \\
$^5$ Observat\'orio Nacional, 20921-400 RJ, Brazil\\
$^6$ {\it Present address:} Osservatorio Astronomico di Brera, Via Brera 28,
Zip Code: 20121, Milano - Italy}
\date{Received April 2002}

\pagerange{\pageref{firstpage}--\pageref{lastpage}}
\pubyear{2002}

\begin{document}

\maketitle

\label{firstpage}

\begin{abstract}

We present a new set of dissipationless N-body simulations aiming to
better understand the pure dynamical aspects of the ``Fundamental
Plane'' (FP) of elliptical galaxies. We have extended our previous
hierarchical merger scheme by considering the Hernquist profile for the
initial galaxy model. Two-component Hernquist galaxy models were also
used to study the effect of massive dark halos on the end-products
characteristics. We have also performed new collapse simulations
including initial spin. We found that the one-component Hernquist
mergers give results similar to those found for the one-component King
models, namely both were able to build-up small scatter FP-like
correlations with slopes consistent with what is found for the
near-infrared FP of nearby galaxies. The two-component models also
reproduce a FP-like correlation, but with a significantly steeper
slope. This is in agreement with what has been found for elliptical
galaxies at higher redshift (0.1 $<$ z $<$ 0.6). We discuss some
structural properties of the simulated galaxies and their ability to
build-up FP-like correlations.  We confirm that collapses generally do
not follow a FP-like correlation regardless of the initial spin. We
suggest that the evolution of gradients in the gravitational field of
the merging galaxies may be the main ingredient dictating the final
non-homology property of the end products.

\end{abstract}

\begin{keywords}
galaxies: elliptical -- galaxies: fundamental parameters --
methods: numerical
\end{keywords}
  
\section{Introduction}

The origin of the ``Fundamental Plane'' relation (hereon FP) of
elliptical galaxies is still unknown, despite of all the efforts to
understand it since it was discovered (\citealt{djo87}, \citealt{dre87}).

On fundamental grounds, the simplest version of the virial theorem
applied to galaxies predicts that they should form a family of objects
following a simple projected relation, involving structural and
kinematical variables, for instance:  $r_e = C_{vir} \sigma_0^2
I_e^{-1}$. In this equation, $C_{vir}$ is a structural-kinematical
parameter, $\sigma_0$ is the central projected velocity dispersion,
$I_e$, the average surface brightness within the effective radius in
linear units, and $r_e$ is the effective radius. The coefficient
$C_{vir}$ relates physical (3-D) to projected variables, like the
velocity dispersion and mass distributions. Hence, $C_{vir} \equiv
C_{vir}(C_r, C_v, M/L)$ depends on kinematical ($C_v$) and structural
($C_r$) coefficients, as well as on the mass-luminosity ratio ($M/L$)
of the systems (c.f.  \citealt{cap95}, hereon CdCC95; \citealt{dan01},
hereon DCdCR01).

We define a family of {\it homologous} galaxies as virialized systems
where the kinematical and structural coefficients are simply constant
for all galaxies, or may change but in a constant ratio throughout the
family.  If furthermore $M/L$ is constant for all galaxies (or
equivalently, $M$ and $L$ may change among them but in a constant
ratio), then $C_{vir}$ is a constant for a given homologous family.

The expression of the FP is similar to that expected from the projected
virial relation but with significantly different exponents and small
scatter throughout:  $r_e \sim \sigma_0^A I_e^B$ (where the exponents
are: $A \sim 1.53$, $B \sim -0.79$, e.g.  \citealt{pah98}).  Thus, in
the case of ellipticals, it is inferred that $C_{vir}$ must vary
monotonically among these galaxies if one is willing to retrieve the
virial relation for these systems.

There are several reasonable alternatives to explain why and how
$C_{vir}$ should vary in order to explain the discrepancy between the
virial theorem and the FP.  One of them assumes a systematic variation
of $M/L$ with the total mass of the system, preserving homology. This
would be responsible for the whole $C_{vir}$ variation (e.g.
\citealt{djo88}, \citealt{djo93}, \citealt{ren93}, \citealt{pah97}).
This dependence would be caused by systematic stellar population (e.g.
mean stellar age, metallicity, etc.) variations with mass.
\cite{pah97} have shown that there is a dependence of the FP tilt with
wavelength, namely $M/L \propto L^{\beta}$ where $\beta$ varies with
the photometric band ($\lambda$) in which the luminosity is measured. This
means that $\beta$ decreases from the $B$ to the $K$ band, although
never reaching the homologous, virial expectations. As discussed by
Pahre \& Djorgovski, the trend of $\beta$ with $\lambda$ cannot be explained
solely by either stellar population models or non-homology (see their
Figure 2).  They conclude that a more complete scenario to explain the
FP tilt has to invoke contributions from both effects. Broken homology
can be achieved both in dissipationless hierarchical merging scenarios
(e.g., CdCC95) and in dissipative mergers of star-forming and gas-rich
spirals, where the roles of star formation histories are emphasized
(c.f.  \citealt{bek98}).  A third line of reasoning for explaining the
FP assumes that a more refined formulation to describe the equilibrium
condition of the luminous component of the elliptical galaxies is
adopting a ``two-component virial theorem'', which assumes of course
that ellipticals are dynamically dominated by a dark halo
(\citealt{dan00}).

In the present work, we study the origin of the FP tilt under the
assumption that elliptical galaxies are more closely described as
non-homologous virialized systems, with $C_v$ and/or $C_r$ varying
monotonically (e.g., CdCC95, \citealt{hjo95},
\citealt{cap97},\citealt{cio96}, \citealt{bus97}, \citealt{gra97},
\citealt{bek98}).  In a hierarchical galaxy formation scenario,
galaxies are built-up by sucessive merge of larger and larger systems.
Recent observations have reinforced the idea of hierarchical merger as
a reasonable mechanism to form elliptical galaxies (e.g.
\citealt{ben98}), although dissipation seems to be an important
ingredient. In any case, numerical investigations of merging seem to be
fundamental in understanding the scaling relations of these objects.
The numerical work of CdCC95 has shown, for instance, that the FP
correlations can arise naturally from objects that are formed by
dissipationless hierarchical mergers of galaxies.  The end products of
their simulations result in a non-homologous family of objects, being
the peculiar non-homology mainly determined by the parameter $C_v$
varying systematically with the initial orbital energy of the galactic
pairs.  In a subsequent investigation, DCdCR01 have shown that
one-component, equal mass collapses of several different initial models
and collapse factors produce nearly homologous families of objects.
This result led DCdCR01 to suggest that the driving mechanism producing
non-homology would be that of merging {\it per se}.

We extend the previous dissipationless numerical
investigations in several aspects. First, the equilibrium models
considered by CdCC95 (King spheres) do not take into consideration a
central density peak.  Recent studies (e.g. \citealt{ger00},
\citealt{sio00}) have demonstrated, however, that the presence of a
central peak (or even the presence of a central black hole) should be
much more common in elliptical galaxies than previously thought. Here
we consider the hierarchical merge of Hernquist models, which present a
central density ``cusp''.  Second, CdCC95 only consider one component
models.  However, it is important to understand the effect of the halo
in the dynamics of merging and the consequences of its influence in the
equilibrium conditions of the whole system (\citealt{dan00}). In this
paper we consider the merge of two-component Hernquist models up to two
generations.

One point not addressed by DCdCR01 was the 
initial difference in spin parameters between proto-galaxies and how that
can introduce non-homology into the structural properties of the final
objects.  In order to address this point, we 
investigate how the spin parameter influences the FP of the collapsed
objects. However, the issue which we leave for future work is the study
of two-component collapses. This is an important problem, since in the
currently accepted cosmological scenarios, the luminous component
collapses in the dark matter halo already virialized some time after
the epoch of equality of matter and radiation energy densities.
Our present goal is to establish the behaviour of only one-component
collapses before analysing two-component ones, which can be studied
under a more general approach as, for instance, drawing the models 
from high resolution cosmological simulations.

This paper is organized in the following way: in Section 2, we
present the simulation setups and initial condition grids; in Section
3, the end products of the simulations are considered in the context of
the FP space and the resulting relaxation histories. Finally, in 
section 4, we discuss our main results.

\section{Simulations Setup and Definition of Characteristic Parameters}

\subsection{Computer Plataforms and Codes}

The simulations were run using two C translations (c.f. \citealt{dub88})
of the TREECODE (\citealt{bar86}): a non-parallel version, which was used
to run less CPU time consuming, one-component model simulations; 
and a parallel version, run for two-component model simulations. 
The computational plataforms used were: (i) For the non-parallel 
code: Workstations Sun-Sparc; Sun-Ultras (1, 2, 5, 10, and 30);
and Sun E250; (ii) For the parallel code: Silicon Graphics Origin 
2000 with four processors using MPI (``Message Passing Interface''), 
IRIX operational system; and a ``cluster'' formed by four Pentium III, 
650 MHz,  working in parallel using LAM (``Local Area Multicomputer'')
6.3.2/MPI 2 C++, Linux.

Quadrupole correction terms, according to \citet{dub88}, were used in
the force calculations for all simulations.  In Table \ref{params} we
list the main parameters of the simulations setup adopted in this
work.  These parameters were carefully chosen in order to conform to
the constraints of resolution and collisionlessness given the total
number of particles used in each type of simulation (more details for
the choice of parameters can be found in CdCC95 and DCdCR01).  In
particular, the choice of the number of particles was also based on the
operational constraints due to CPU times. Merging generaly involves
CPU time-consuming runs for it includes the evolution since the
initial orbital phase, before the effective merge of the systems.  This
forced us to use a relatively small number of particles to cover a
wider grid of initial conditions.  These numbers, however, are
well above the lower bound discussed in DCdCR01.

\begin{table}
\caption{Initial Parameters of the Codes}
\label{params}
\vspace{0.3cm}
\begin{tabular}{ll} \hline
\multicolumn{1}{c} {Parameter} & \multicolumn{1}{c} {Value} \\ 
\hline \hline
$\theta$: tolerance   & $0.8$ \\ \hline
$\epsilon$: softening   &      \\
$~~~\rightarrow$ Non-parallel code & $0.05$ \\
$~~~\rightarrow$ Parallel code &  \\
$~~~~~~~~~~~~~~$ Luminous component & $0.07$ \\
$~~~~~~~~~~~~~~$ Dark component  & $0.7$ \\ \hline
$\Delta t$: time integration step & $0.025$ \\ \hline
\end{tabular}
\end{table}

\subsection{Initial Condition Grid of the Models}

\subsubsection{Computational Units}

The units used in our simulations were all set to match those of CdCC95
and DCdCR01:  the mass and lenght units were set to $M_U \equiv
10^{10}$ $M_{\odot}$ and $L_U \equiv 1$ kpc, respectively.  These
values, and $G \equiv 1$, fix our time and velocity units to $T_U
\equiv 4.72$ Myr and $v_U \equiv 207$ km s$^{-1}$, respectively.

\subsubsection{The Merger Models}

The initial equilibrium models for the merger pair were each obtained
from $N$ particles random realizations of spheres in hydrostatic
equilibrium, obeying the Hernquist profile (c.f. \citealt{her90}). We
considered both one as well as two component models, in equilibrium in
the common potential. 

The reasons for the choice of a Hernquist profile for the luminous and
dark components were based on the desire to test whether models
including a central density 'cusp' (in this case, the Hernquist models
provided us with this characteristic) could also reproduce the results
by CdCC95. Since the FP parameters refer to central (effective)
quantities, the idea was to test whether the results changed sensibly
or not with the inclusion of an initial 'cusp' in the models.  In
particular, the reasons for the adoption of the Hernquist profile also
for the halo (instead of, e.g., a truncated isothermal profile, e.g.
\cite{wal96}) comes from the fact that the density profile behaves as
$\sim r^{-1}$ at small radii resembling the Navarro, Frenk \& White
(1997) `universal' profile, which results from cosmological
simulations.

The Hernquist models were truncated at a specified energy cut-off:
$10\%$ least bound particles of the model were eliminated. Hence, the
original Hernquist one-component model had a mass of $1 M_U$. The
truncated model resulted in a mass of $0.9 M_U$.  The same was applied
to the two-component models, where both the luminous as well as the
dark component were truncated by the same factor ($10\%$): the luminous
component has a mass of $0.9 M_U$ and the dark component, a mass of $9
M_U$.

We assigned to the one-component Hernquist mergers the labels: D, E e
F, according to which generation they belong (D: first, E: second, and
F: third).  The total initial values of mass and number of particles of
the D mergers were: $M_{tot, D} = 1.8 ~M_U$ and  $N_{tot, D} = 8194$,
respectively (these values refer to the sum of the two initial merging
models, not to one model alone).

The two-component mergers were assigned with labels Z (Z01-Z09: first
generation; Z10-Z13: second generation).  We chose the initial luminous
($M_L$) to dark ($M_H$) mass ratio of the initial two-component
Hernquist model as $\mu_{init} \equiv M_L/M_H = 0.1$ (the results of
\cite{mih98} favor $M_{H} \sim (4-8)~ M_{disk+bulge}$ for NGC 7252,
suggesting our mass ratio is reasonable). The total initial mass of the
Z mergers was $M_{tot, Z} = 19.8 ~M_U$, with a total of $N_{tot, Z} =
9000$ particles.  Each initial two-component Hernquist model therefore
has a luminous mass of $0.9$ ($2250$ particles) and a dark mass of $9$
(also $2250$ particles).  Note that since the number of particles per
component is the same, the mass per dark matter particle is greater
than that of the luminous particle by a factor $10$.

The initial ratio of the effective (half-mass) radius of dark matter to
that of the luminous component was $a_{H} = 10 ~a_{L}$.  Here we
briefly discuss the reasoning for choosing these ratios.  \cite{sal00}
find for disk galaxies $r_0 \sim (4-7) R_d$, where $r_0$ is the halo
core radius of the \cite{bur95} model ($r_0$ is of same order as $r_c$,
the core radius of the modified isothermal model). $R_d$ is the disk
scale radius.  Noticing that the effective radius for spirals, $\langle
R^S_e \rangle$, is approximately related to $R_d$ by $\langle R^S_e
\rangle \approx 1.2 R_d$, then $r_0 \approx (3.3-5.8) \langle R^S_e
\rangle$.  Noticing also that $\langle R^S_e \rangle \approx R_e^E$ for
$L=L_*$ galaxies, where $R_e^E$ is the effective radius for giant
ellipticals, and that the $R_e \approx 1.8153 ~a_{L}$ (c.f.
\citealt{her90}), where $a_{L}$ is the scale radius of the Hernquist
profile for the luminous component, one can infer that the results of
Salucci \& Burkert imply $r_0 \approx (6-10) a_{L}$.  Assuming $a_0$ is
of the same order as $a_{H}$, the scale radius of the Hernquist profile
for the dark matter component, there is a compatibility between our
adopted values for the initial ratio ($a_{H} = 10 ~a_{L}$) and the
results by Salucci \& Burkert (although their analysis was based on
spiral galaxies).  \cite{ger01}, on the other hand, find that $r_{c,h}
\approx 1.2 ~R_e$ for E0 ellipticals, where $r_{c,h}$ is the ``minimum
halo model'' core radius (c.f. \citealt{kro00}). Again, this can be
translated to $r_{c,h} \approx 2.2 ~a_{L}$.  It is not at all clear the
correspondence between $r_{c,h} $ and $a_{H}$, but if they have the
same order of magnitude, it would seem to imply our value ($a_{H} = 10
~a_{L}$) would be somewhat higher than adequate.  On the other hand,
however, there are some works on the morphology and kinematics of tidal
tails of merger models, where some inferences can be made on the halo
properties by a comparison with simulations. \cite{mih98}, for
instance, study models with ratios of mass and radius within the range
of our model.  They find a good fit to NGC 7252, favoring relatively
compact, low-mass halos for the progenitors of the merger.  Although
their results are somewhat idealistic, our models do not seem to be
imcompatible with what is usually adopted in the literature. However,
in face of the uncertainties for a reasonable value for the effective
(half-mass) radius of dark matter to that of the luminous component, we
check the dependency of the FP results on the choice of this ratio. To
that end, have run two sets of nine simulations similar to the Z
models, but using a more compact halo, namely: $a_{H} = 3 ~a_{L}$ and
$a_{H} = 5 ~a_{L}$. These models are labeled Z01b-Z09b and Z01c-Z09c,
respectively.

The initial merging conditions were characterized according to a
generalization of a prescription described in \citet{bin87} (c.f.
CdCC95).  In this formulation, the initial orbit of the binary galaxy
system is characterized, essentially, by the energy and angular
momentum of the Keplerian orbit of two point masses equivalent to the
initial galaxies. We defined the dimensionless energy and angular
momentum of the orbit as:
\begin{equation}
\hat{E} \equiv {E_{orb} \over {1 \over 2} \mu \overline{\langle v^2 \rangle}}
\end{equation}
\begin{equation}
\hat{L} \equiv {L_{orb} \over \mu r_h \overline{\langle v^2 \rangle}^{1 \over 2}} 
\end{equation}
with $\overline{\langle v^2 \rangle} \equiv 
\sqrt{\langle v^2_1 \rangle  \langle v^2_2 \rangle}$, 
$\overline{r}_h \equiv \sqrt{r_{h1} r_{h2}}$, where $r_{hi}$ ($i=1,2$)
is the half-mass radius of the system $i$, and $\mu$ is the reduced
mass of the system. A third parameter
depends only on the dynamical structure of the initial galaxies:
\begin{equation} 
A \equiv {2 G M \over \overline{r}_h
\overline{\langle v^2 \rangle}}, 
\end{equation} 
which presents a not very large variation 
(${~}^{<}_{\sim}~ 20 \%$) among the initial models ($A \sim 17$).

The initial separation of the models was chosen as $\sim 4 r_h$ for the
parabolic and hyperbolic orbits, and the apocenter position for the
closed orbits.  These initial separations were chosen considering that
they should not be too close (implying that tidal effects would be
artificially disregarded due to the spherical symmetry of the initial
models) nor too far away, so that time consuming CPU runs were
avoided.

By using this grid of initial conditions, the models merged and evolved
up to $\sim 30$ ``crossing times'' ($T_{cr} = GM^{5/2}/(2|E|)^{3/2}$),
when quantities like half-mass radius ($r_h$) and virial ratio
($\beta_v \equiv {2K \over |W|}$) indicated no significant variation of
the resulting system ($\Delta r_h/r_h {~}^{<}_{\sim} ~0.5 \%$, and
$\Delta \beta_v/\beta_v {~}^{<}_{\sim} ~ 1 \%$ after $\sim 10
~T_{cr}$).

In Tables \ref{tab_Hmergers} (one-component models) and
\ref{tab_H2Cmergers} (two-component models) we list the initial
condition grids of the merger simulations.  The two-component merger
simulations using a more compact halo than the Z models (i.e., the Zb
and Zc models, were $a_{H} = 3 ~a_{L}$ and $a_{H} = 5 ~a_{L}$,
respectively) are also included in Table \ref{tab_H2Cmergers}. We also
list in Table \ref{tab_Kmergers} the simulations performed by CdCC95,
including several third generation simulations not previously published
(the total number of particles for the first generation of these
mergers is 8192).

\subsubsection{The Collapse Models}

In Table \ref{tab_colapsos} we include the simulations
performed by DCdCR01 for easy reference. Details of the collapse models
can be found in DCdCR01. Here we give a brief summary of these collapse
simulations:  three different initial collapse models were used
(labeled K, A and C).  All the initial models have total mass $M = 20
~M_U$ and radius $R = 20 ~L_U$, except the C models, which have $R=100
~L_U$. The K models were constructed from 8192 Monte Carlo realizations
of a spherical isotropic King model.  The A models were constructed
from spherical $r^{-1}$ models of 16384 Monte Carlo particle
realizations (\citealt{agu90}). The C models were constructed according to
\citet{car95}, with 4096 particles.  The initial velocities of these
models had gaussian profiles.   All models were pertubed according to
the collapse factor parameter $\beta$  ($0 ~{~}^{<}_{\sim}~ \beta
~{~}^{<}_{\sim}~ 1$, where $\beta \equiv 2K_0/|W_0|$; $K_0$ is the
initial kinetic energy and $W_0$, the initial potential energy of the
system).  To the C models, a Hubble flow assuming $H_0 = 65$
km~s$^{-1}$~Mpc$^{-1}$ has been incorporated. We generically denoted
``cold'' collapses those resulted from $\beta \rightarrow 0$, and
``hot'' collapses those resulted from $\beta \rightarrow 1$.

We have included also two sets of collapse simulations with a range
of initial solid body rotation, not discussed in DCdCR01. 
We have included spins to the unperturbed, initial A model in order to study
their effects in the final systems. The reason to
focus on the A models is because these collapses spread in the FP
space, contrary to the K models.  Although the C models are more
`realistic' (they evolve from small pertubations in a Hubble
expansion), because of being dinamically more complex we have avoided
them our analysis of the spin effect on the FP (see details in
DCdCR01).

The method we assume here is inspired on that of \cite{wil82}.  We have
given a solid body angular velocity, $\omega$, to each particle of the
unperturbed A model. The value of $\omega$ was chosen such that the resulting
total kinetic energy after including the spin was a fraction $\gamma$
greater than the initial total kinetic energy (i.e., without the
rotational motion). In other words, $\gamma = |K_{prog} -
K_{prog,spin}|/K_{prog}$. For the first set of collapse simulations
with spin, which we label AS1 models, the rotational perturbation
chosen was small, $\gamma \sim 5\%$.  The total velocity squared of each
particle was then reduced by a range of $\beta$ factors, producing 9
spin models with different collapse fators. These collapses can be
directly compared to the A collapses studied by DCdCR01.  For the
second set of models, $\gamma$ was chosen in order to impose a maximum
perturbation to the A progenitor such that, after reducing the velocity
field by the ``hottest'' perturbation we are considering (viz. $\log
\beta=-0.01$), the resulting model was barely able to collapse (total
binding energy was $\sim -0.007$). The value the perturbation in
this case was $\gamma \sim 38\%$. The perturbed progenitor was ``cooled''
by the same $\beta$ factors as the AS1 models.  This second set of
models was labeled AS3 models.  These new collapse simulations are
listed in Table \ref{tab_colapsos_spin}.

Note that the structure (viz. potential energy) of the A models used
here to construct the spin models did not allow the inclusion of a
higher initial spin than that of the AS3-09 ($\log \beta=-0.01$) model
without disrupting the system (viz., expanding it instead of making it
collapse). Higher spin rates could have been used, but that would imply
changing the structure of the progenitor by, e.g., reconfiguring the
positions of the particles (viz., by decreasing the gravitational
radius of the system) or incrasing the total mass. That would change
considerably the structure of the progenitor and would not allow a
direct comparison with the collapse models of DCdCR01. Hence, these
collapses with spin just represent models were an initial rotational
``perturbation'' was applied to the system. This alowed us to
keep  the same initial structure of the progenitor of the A models,
used by DCdCR01, and still make the resulting model collapse according
to the $\beta$ factor.

\begin{table*}
\caption{Mergers of Hernquist Models (one-component)} 
\label{tab_Hmergers}
\vspace {0.3cm}
\scriptsize
\begin{tabular}{ccccccccccccccccccc}
\multicolumn{6}{c}{!st. Generation} & \multicolumn{7}{c}{2nd. Generation} & \multicolumn{6}{c}{3rd. Generation} \\ \hline
Run  & $\hat{E}$  & $\hat{L}$   & Sep. & $N_{part}$ &  & Run & $\hat{E}$ & $\hat{L}$ & Sep. & $N_{part}$ & Progen. &  &  Run & $\hat{E}$ & $\hat{L}$ & Sep. & $N_{part}$ & Progen.  \\ \hline \hline
D01  &  0.0   & 0 & 15.0 & 8194 &  & E1 &  0.0  & 0 & 16.0 & 15777 & D01-D02  &  & F01  & 0.0   & 0 & 22.0 & 31260 & E02-E02   \\
D02  & -3.0   & 1 & 11.2 & 8194 &  & E2 & -2.0  & 1 & 46.4 & 16064 & D03-D04  &  & F02  & -1.0  & 1 & 25.0 & 30209 & E02-E05  \\
D03  & -1.0   & 1 & 33.8 & 8194 &  & E3 & -1.0  & 1 & 89.1 & 16248 & D07-D08  &  & F03  & -10.0 & 1 & 15.2 & 28497 & E02-E04   \\
D04  & -7.5   & 2 &  4.0 & 8194 &  & E4 &  0.0  & 2 & 12.0 & 15416 & D01-D01  &  & F04  & -7.0  & 2 & 21.8 & 27446 & E04-E05  \\
D05  & -1.0   & 2 & 33.4 & 8194 &  & E5 & -3.0  & 0 & 12.0 & 15766 & D03-D03  &  &     &       &   &      &       &  \\
D06  &  0.5   & 2 & 15.0 & 8194 &  & E6 & -0.3  & 0 & 100.0 & 15523 & D05-D06 &  &     &       &   &      &       & \\
D07  & -6.9   & 3 &  3.4 & 8194 &  &    &      &   &       &      &          &   &    &       &   &      &       &  \\
D08  & -2.8   & 3 & 10.9 & 8194 &  &    &      &   &       &      &          &   &    &       &   &      &       & \\
D09  &  0.0   & 3 & 15.0 & 8194 &  &    &      &   &       &      &          &   &     &       &   &      &       &   \\
D10  & -5.0   & 0 & 10.0 & 8194 &  &    &      &   &       &      &          &   &     &       &   &      &       &   \\ \hline
\end{tabular}
\end{table*}

\begin{table*}
\caption{Mergers of Hernquist Models (two-components)} 
\label{tab_H2Cmergers}
\vspace {0.3cm}
\scriptsize
\begin{tabular}{cccccccccccc}
\multicolumn{6}{c}{1st. Generation} & \multicolumn{6}{c}{2nd. Generation} \\ \hline
Run   & $\hat{E}$  & $\hat{L}$  & Sep. & $N_{part}$ & & Run & $\hat{E}$ & $\hat{L}$ & Sep. & $N_{part}$ & Progen. \\ \hline \hline
Z01   & -4.0   & 0 & 70.0     & 9000 &  & Z10 & -2  & 1 & 207.66  & 17807 & Z06-Z07    \\
Z02   & -4.0   & 1 & 70.0     & 9000 &  & Z11 & -1  & 1 & 416.04  & 17507 & Z09-Z02    \\
Z03   & -3.0   & 0 & 70.0     & 9000 &  & Z12 & -1  & 1 & 416.04  & 17944 & Z01-Z01    \\
Z04   & -3.0   & 1 & 138.2    & 9000 &  & Z13 &  0  & 0 &  70.00  & 17954 & Z01-Z04    \\
Z05   & -2.0   & 0 & 70.0     & 9000 &  &     &     &   &         &       &            \\
Z06   & -2.0   & 1 & 207.6    & 9000 &  &     &     &   &         &       &            \\
Z07   & -2.0   & 2 & 205.4    & 9000 &  &     &     &   &         &       &            \\
Z08   &  0.0   & 0 & 70.0     & 9000 &  &     &     &   &         &       &            \\
Z09   &  0.5   & 0 & 70.0     & 9000 &  &     &     &   &         &       &            \\ \hline
Z01b  & -4.0   & 0 & 70.0     & 9000 &  &     &     &   &         &       &            \\
Z02b  & -4.0   & 1 & 70.0     & 9000 &  &     &     &   &         &       &            \\
Z03b  & -3.0   & 0 & 70.0     & 9000 &  &     &     &   &         &       &            \\
Z04b  & -3.0   & 1 & 138.2    & 9000 &  &     &     &   &         &       &            \\
Z05b  & -2.0   & 0 & 70.0     & 9000 &  &     &     &   &         &       &            \\
Z06b  & -2.0   & 1 & 207.6    & 9000 &  &     &     &   &         &       &            \\
Z07b  & -2.0   & 2 & 205.4    & 9000 &  &     &     &   &         &       &            \\
Z08b  &  0.0   & 0 & 70.0     & 9000 &  &     &     &   &         &       &            \\
Z09b  &  0.5   & 0 & 70.0     & 9000 &  &     &     &   &         &       &            \\ \hline
Z01c  & -4.0   & 0 & 70.0     & 9000 &  &     &     &   &         &       &            \\
Z02c  & -4.0   & 1 & 70.0     & 9000 &  &     &     &   &         &       &            \\
Z03c  & -3.0   & 0 & 70.0     & 9000 &  &     &     &   &         &       &            \\
Z04c  & -3.0   & 1 & 138.2    & 9000 &  &     &     &   &         &       &            \\
Z05c  & -2.0   & 0 & 70.0     & 9000 &  &     &     &   &         &       &            \\
Z06c  & -2.0   & 1 & 207.6    & 9000 &  &     &     &   &         &       &            \\
Z07c  & -2.0   & 2 & 205.4    & 9000 &  &     &     &   &         &       &            \\
Z08c  &  0.0   & 0 & 70.0     & 9000 &  &     &     &   &         &       &            \\
Z09c  &  0.5   & 0 & 70.0     & 9000 &  &     &     &   &         &       &            \\ \hline
\end{tabular}
\end{table*}

\begin{table*}
\caption{Mergers of King Models (one-component, CdCC95 plus 3rd. generation new data)} \label{tab_Kmergers}
\vspace {0.3cm}
\scriptsize
\begin{tabular}{ccccccccccccc} 
\multicolumn{4}{c}{1st. Generation} & \multicolumn{5}{c}{2nd. Generation} & \multicolumn{4}{c}{3rd. Generation} \\ \hline
Run & $\hat{E}$ & $\hat{L}$  & & Run & $\hat{E}$ & $\hat{L}$  & Progen. &  & Run & $\hat{E}$ & $\hat{L}$  & Progen.  \\ \hline \hline
R1 &  0.0   & 0 &  & H1  &  0.5 & 3   & R17-R17 &  & H14  & -2.0  & 3   & R6-H3  \\
R2 & -4.0   & 1 &  & H2  & -2.0 & 1   & R6-R6   &  & H15  & -2.0  & 4   & H1-H3 \\
R3 & -3.0   & 1 &  & H3  & -4.0 & 1   & R17-R17 &  & H16  & 0.5   & 4   & H1-H1 \\
R4 & -2.0   & 1 &  & H4  & -2.0 & 1   & R8-R8   &  & H17  & -1.77 & 4   & H13-H13 \\
R5 & -1.0   & 1 &  & H5  & -2.0 & 1   & R14-R14 &  & H18  & -3.0  & 3   & H13-H13 \\
R6 & 0.5    & 1 &  & H6  & -2.0 & 3   & R2-R2   &  & H19  & -0.5  & 3   & H10-H19 \\
R7 & -7.5   & 2 &  & H7  & -2.0 & -3  & R2-R2   &  &      &       & &  \\
R8 & -5.7   & 2 &  & H8  & -2.0 & 3   & R9-R9   &  &      &       & &  \\
R9 & -1.0   & 2 &  & H9  & -3.0 & 2   & R9-R9   &  &      &       & &   \\
R10 & 0.0   & 2 &  & H10 & -3.0 & -2  & R9-R9   &  &      &       & &  \\
R11 & 0.5   & 2 &  & H11 & -2.0 & 1   & R10-R10 &  &      &       & &  \\
R12 & -7.9  & 3 &  & H12 & -2.0 & 1   & R1-R1   &  &      &       & &   \\
R13 & -6.9  & 3 &  & H13 &  0.5 & 2   & R11-R11 &  &      &       & &   \\
R14 & -5.1  & 3 &  &     &      &     &         &  &      &       & &    \\
R15 & -2.8  & 3 &  &     &      &     &         &  &      &       & &    \\
R16 & -1.0  & 3 &  &     &      &     &         &  &      &       & &    \\
R17 & 0.0   & 3 &  &     &      &     &         &  &      &       & &    \\ \hline 
\end{tabular}
\end{table*}

\begin{table*}
\caption{Collapses (one-component, DCdCR01)} \label{tab_colapsos}
\vspace {0.3cm}
\scriptsize
\begin{tabular}{cccccccccccccc} \hline
\multicolumn{2}{c}{K Models} & \multicolumn{1}{c}{} & \multicolumn{2}{c}{A Models} & \multicolumn{1}{c}{} & \multicolumn{8}{c}{C Models} \\
\cline{1-2} \cline{4-5} \cline{7-14} \\
\multicolumn{2}{c}{$N_{part} = 8192$} & \multicolumn{1}{c}{} & \multicolumn{2}{c}{$N_{part} = 16384$} & \multicolumn{1}{c}{} & \multicolumn{8}{c}{$N_{part} = 4096$} \\
\multicolumn{2}{c}{} & \multicolumn{1}{c}{} & \multicolumn{2}{c}{} &  \multicolumn{1}{c}{} & \multicolumn{2}{c}{$n=1$} & \multicolumn{1}{c}{} & \multicolumn{2}{c}{$n=0$} & \multicolumn{1}{c}{} & \multicolumn{2}{c}{$n=2$} \\ \hline \hline
$\log \beta$ & Run & & $\log \beta$ & Run & & $\log \beta$ & Run & & $\log \beta$ & Run & & $\log \beta$ & Run  \\ \hline
-4.00 & K01  & & -4.00  &  A01   & & -3.75 & C01  & & -3.75 & C11  &  & -3.75 & C21   \\
-3.75 & K02  & & -3.50  &  A02   & & -3.50 & C02  & & -3.50 & C12  &  & -3.50 & C22   \\
-3.50 & K03  & & -3.00  &  A03   & & -3.25 & C03  & & -3.25 & C13  &  & 3.25  & C23   \\
-3.25 & K04  & & -2.50  &  A04   & & -3.00 & C04  & & -3.00 & C14  &  & -3.00 & C24   \\
-3.00 & K05  & & -2.00  &  A05   & & -2.50 & C05  & & -2.50 & C15  &  & -2.50 & C25   \\
-2.75 & K06  & & -1.50  &  A06   & & -2.00 & C06  & & -2.00 & C16  &  & -2.00 & C26   \\
-2.50 & K07  & & -1.25  &  A07   & & -1.50 & C07  & & -1.50 & C17  &  & -1.50 & C27   \\
-2.25 & K08  & & -1.00  &  A08   & & -1.00 & C08  & & -1.00 & C18  &  & -1.00 & C28   \\
-2.00 & K09  & & -0.75  &  A09   & & -0.90 & C09  & & -0.90 & C19  &  & -0.90 & C29   \\
-1.75 & K10  & & -0.50  &  A10   & & -0.80 & C10  & & -0.80 & C20  &  & -0.80 & C30   \\
-1.50 & K11  & & -4.10  &  A01b  & & -4.00 & C01b & &       &      &  &       &           \\
-1.25 & K12  & & -3.60  &  A02b  & & -3.60 & C02b & &       &      &  &       &             \\
-1.00 & K13  & & -3.40  &  A03b  & & -3.40 & C03b & &       &      &  &       &            \\
-0.75 & K14  & & -3.10  &  A04b  & & -3.10 & C04b & &       &      &  &       &            \\
-0.50 & K15  & & -2.75  &  A05b  & & -2.25 & C06b & &       &      &  &       &            \\
-0.25 & K16  & & -2.25  &  A06b  & & -1.75 & C07b & &       &      &  &       &            \\
-0.01 & K17  & & -1.75  &  A07b  & & -1.25 & C08b & &       &      &  &       &          \\ 
      &      & & -1.25  &  A08b  & & -0.25 & C09b & &       &      &  &       &          \\ 
      &      & & -0.95  &  A09b  & & -0.10 & C10b & &       &      &  &       &          \\ 
      &      & & -0.85  &  A10b  & &       &      & &       &      &  &       &          \\ 
      &      & & -0.25  &  A11   & &       &      & &       &      &  &       &          \\ 
      &      & & -0.10  &  A12   & &       &      & &       &      &  &       &          \\

\hline
\end{tabular}
\end{table*}

\begin{table*}
\caption{Collapses (the ``A'' Model Progenitor with Initial Spin)} \label{tab_colapsos_spin}
\vspace {0.3cm}
\scriptsize
\begin{tabular}{ccccc} \hline
\multicolumn{2}{c}{AS1 Models} & \multicolumn{1}{c}{} & \multicolumn{2}{c}{AS3 Models} \\
\multicolumn{2}{c}{$N_{part} = 16384$, $\gamma = 5\%$} & \multicolumn{1}{c}{} & \multicolumn{2}{c}{$N_{part} = 16384$, $\gamma = 38\%$} \\ \hline \hline
$\log \beta$ & Run & & $\log \beta$ & Run \\ \hline
-4.00 & AS1-01  & & -4.00  &  AS3-01    \\
-3.50 & AS1-02  & & -3.50  &  AS3-02  \\
-3.00 & AS1-03  & & -3.00  &  AS3-03  \\
-2.50 & AS1-04  & & -2.50  &  AS3-04    \\
-2.00 & AS1-05  & & -2.00  &  AS3-05    \\
-1.50 & AS1-06  & & -1.50  &  AS3-06   \\
-1.00 & AS1-07  & & -1.00  &  AS3-07   \\
-0.50 & AS1-08  & & -0.50  &  AS3-08   \\ 
-0.01 & AS1-09  & & -0.01  &  AS3-09  \\

\hline
\end{tabular}
\end{table*}

\section{The Fundamental Plane of End Products}

\subsection{The FP space}

\begin{figure}
\epsfig{file=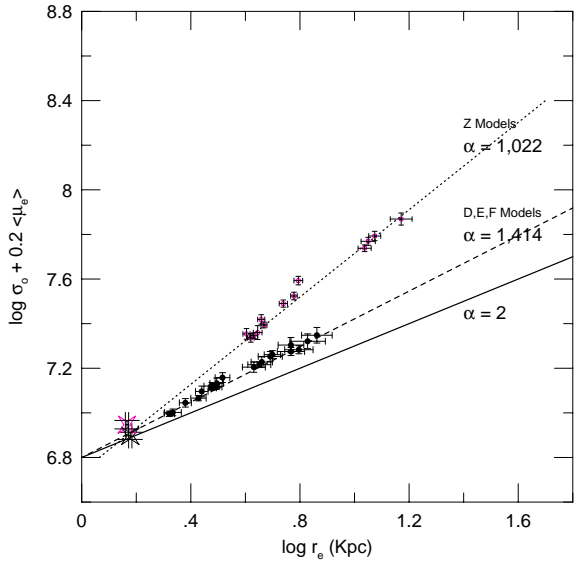,width=8cm}
\caption{Results of the merger simulations in terms of the FP
paremeters.  In the case of two-component models, only the luminous
component is presented. The stellar symbol represents the progenitor.
The continuous line represent the prediction of the virial theorem for
homologous systems. \label{fp-lum}} 
\end{figure}

We follow the method given by CdCC95 to compute the characteristic FP
variables ($r_e$, $\mu_e \equiv -2.5 \log I_e$ and $\sigma_0$) of the
simulated models.  The variables $\sigma_0$ and $\mu_e$ were combined
in the vertical axis according to the usual representation of the FP
projected onto the cartesian plane $\log{r_e} \times \log{\sigma_o} + 
\beta \mu_e$. In all the cases the 3-variate best-fit solutions for a plane 
gave $\beta = \rm 0.2$ to within 10\%, so we decided to keep this coefficient
fixed at 0.2 in order to find the orthogonal least squares solutions for the 
other coefficients, viz. the slope $\alpha$ (the FP `tilt'), and the 
intercept of the fitted plane. 

Before analysing the final simulated models in the FP space, however,
it is worth to comment that they reproduce the general structural
characteristics of elliptical galaxies, e.g. projected triaxialities
(from E0 to E5 elliptical objects) and surface density profiles
(following the Sersic law).  A detailed discussion on the structural
properties of the simulated models is given in \citet{dantastese}.

In Fig. \ref{fp-lum} we present the characteristic FP parameters of the
objects resulting from the merging of one (D, E e F mergers) and two (Z
mergers) component models. In the case of two-component models, the
data shown in this figure are relative to the luminous component.  The
best fit values of the FP slope ($\alpha$) found for these
simulations are indicated in the figures. The continuous line ($\alpha
= 2$) represent the prediction of the virial theorem for homologous,
constant $M/L$, systems. In Table \ref{tab_ajus_fus}, we present the
results of the best fit values of the data here discussed as well as
the results obtained by CdCC95 and DCRdC01.  The results indicate that
one-component Hernquist mergers (D, E, F) also reproduce reasonably the
FP tilt of the elliptical galaxies, consistent with the results
obtained with the King models of CdCC95. That is, for both cases, the FP
slopes are consistent, within the errors, to that observed for infrared
FP of nearby galaxies, that is, $\alpha = 1.53 \pm 0.08$
(\citealt{pah98}).

\begin{figure}
\epsfig{file=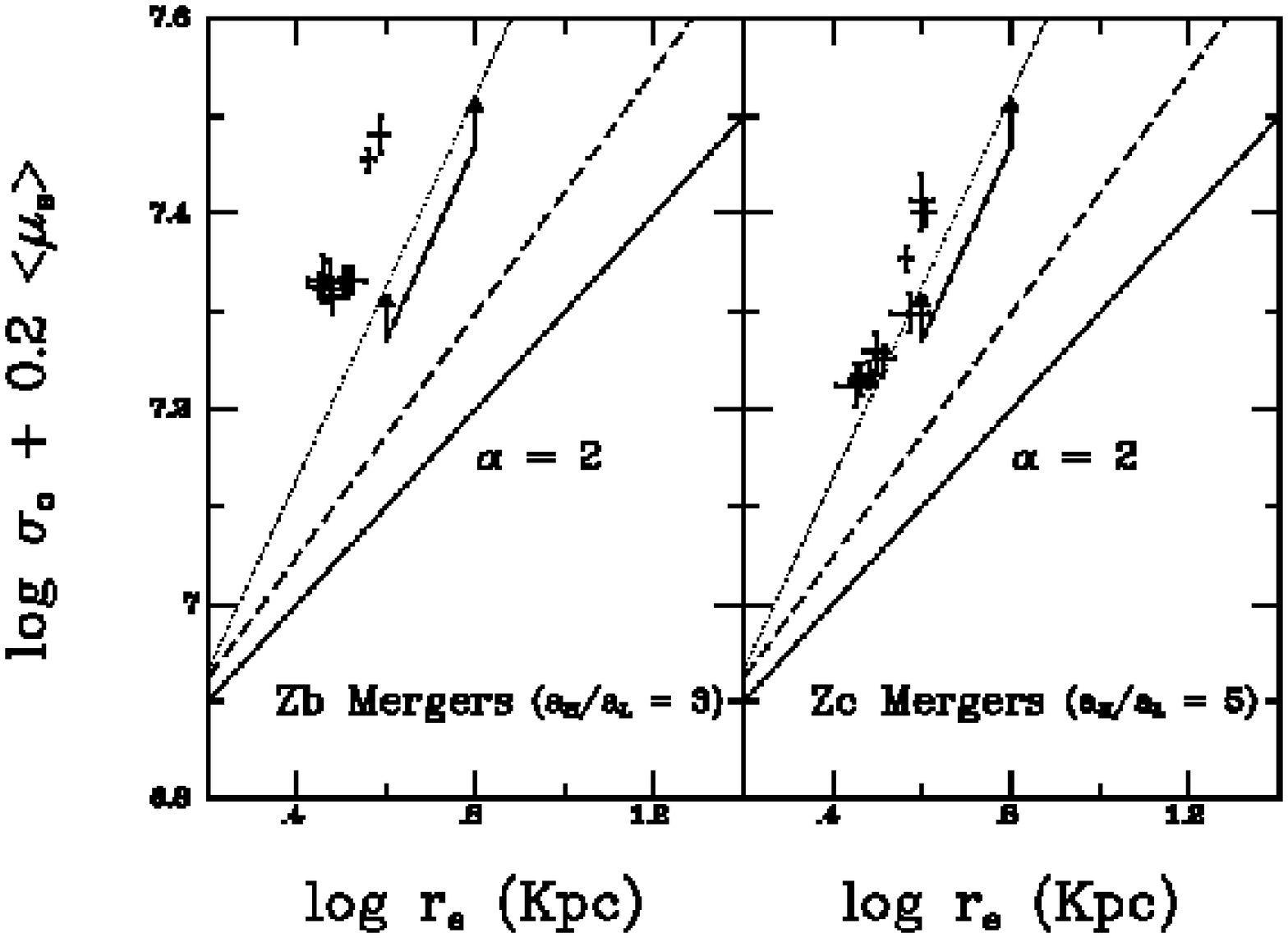,width=8cm}
\caption{Results for the luminous component of 
the Hernquist merger simulations, using
more compact halos.  {\it Left panel:} Zb mergers {\it Right panel:} Zc
mergers.  The three lines on these panels are reproductions of the fits
shown in Fig.1, for comparison. In particular, the solid line is the
prediction of the virial theorem for homologous systems.  The arrows
over the dotted line represent the range occupied by the first
generation Z models ($a_H/a_L = 10$), for comparison.
\label{fp-zbc}}
\end{figure}

On the other hand, the luminous/barionic objects resulting from the the
two-component mergers form a family with a steeper relation in
comparison to one-component mergers. (It is interesting to note that if
we consider an equivalent to the FP space but for the dark halos of
these merger remnants, we find that they constitute an approximate
homologous family of objects, as indicates the value $\alpha = 1.872$,
in Table \ref{tab_ajus_fus}.) In order to test the effects of a more
compact halo on these results, we ran two groups of two-component
Hernquist mergers with different ratios for the halo to luminous radius
(as discussed in Section 2.2.2). Unlike the Z models, these runs were
not followed for subsequent (viz., second, etc) generations because of
CPU time limitations.  In Fig. \ref{fp-zbc}, we show how these more
compact halo mergers distribute in the FP space.  The arrows over the
dotted lines in that figure represent the range occupied by the first
generation Z models ($a_H/a_L = 10$), for comparison.  It is
interesting to notice that the luminous component of the most compact
halo models (Zb models), with most negative $E_{orb}$'s, tend to
cluster in the FP space in a similar manner as the K collapses. All
other models tend to spread out sensibly along a FP-like relation.
Also, the luminous body of these models (Zb and ZC models) tend to
settle into systematically lower values of $r_e$ and at higher values of
$y$ ($\equiv \log \sigma_0 + 0.2 \langle \mu_e \rangle$) than the Z
models ($a_H/a_L = 10$). The FP `tilts' of these models
suggest a marginally steeper `tilt' than the Z models.

\begin{figure}
\epsfig{file=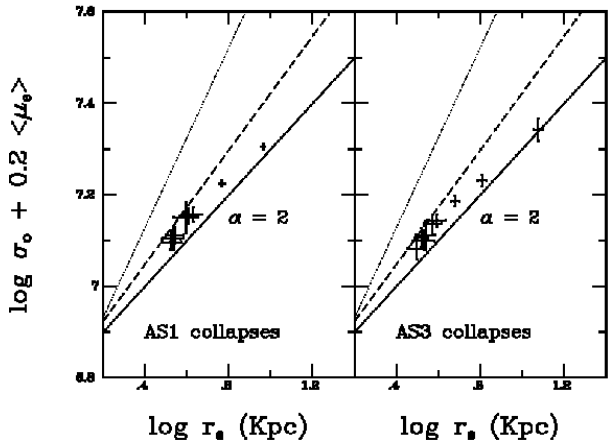,width=8cm}
\caption{Results of the collapse simulations with initial spin in
terms of the FP paremeters. 
{\it Left panel:} AS1 collapses, constructed from the 
progenitor of the ``A'' models. This progenitor received a solid body 
rotation resulting on a $\sim 5\%$ perturbation to the
initial total kinetic energy of the system, and then ``cooled''
by a range of $\beta$ factors. {\it Right panel:} AS3 models,
produced in similar manner as the AS1 collapses, but receiving
a larger initial rotational perturbation ($\sim 38\%$).
The three lines on these panels are reproductions of the fits 
shown in Fig.1, for comparison. In particular, the solid line
is the prediction of the virial theorem for
homologous systems. \label{fp-spin}}
\end{figure}

As already mentioned, we have performed two groups (AS1 and AS3 models)
of collapse simulations with initial spin in order to verify the effects 
of the inclusion of rotation on the results by DCdCR01, where
evidences for homology were found for pure collapses. The resulting
FP `tilts' for both groups ($\alpha_{AS1} = 2.204 \pm  0.158$; $\alpha_{AS3}
= 2.306 \pm 0.250$) suggests that the resulting models are slightly
non-homologous, but in the {\it opposite} (viz. $\alpha > 2$)
sense from the observed FP `tilt' of elliptical galaxies (c.f. Fig.
\ref{fp-spin}).

All these new collapse models evolved for more than $2$ Gyr ($\sim 30
T_{cr}$), however, the ``hottest'' models (viz., AS1-09 and AS3-09,
both with initial $\log 2K/|W| = -0.01$) still presented a virial ratio
oscillating around $2K/W \sim -1.4$ by that time. These ``hottest''
models seem to evolve very slowly and still did not reach complete
virial equilibrium after $2$ Gyr, whereas all the other models were
already well virialized.  Removing these ``hottest'' collapse models
results in $\alpha_{AS1-0.01} = 1.966 \pm 0.270$ and $\alpha_{AS3-0.01}
= 2.190 \pm 0.349$, which are compatible with homology.

Although we cover a reasonable range of $\beta$'s, it is not clear
whether the inclusion of more intermediate $\beta$ collapses would
necessarily improve the statistics (i.e., decrease the error bar of the
fit), since the ``coldest'' models tend to cluster in the FP space.  On
the other hand, the inclusion of ``hotter'' collapses would only
exacerbate the observed ``inverted'' (viz. $\alpha > 2$) non-homology.
These results seems to indicate that an initial spin is not sufficient
to produce non-homology, at least of the same nature of mergers.  

\begin{table}
\caption{PF Best Fit Values} \label{tab_ajus_fus}
\vspace {0.3cm}
\scriptsize
\begin{tabular}{lll} 
\multicolumn{1}{c} {Model} & \multicolumn{1}{l} {$\alpha \pm \delta\alpha$} & \multicolumn{1}{l} {$N_{fit}$}\\ 
\hline \hline 
\multicolumn{3}{c} {One-component Models:} \\ \hline \hline
D, E, F Mergers                          & $\alpha = 1.414 \pm 0.132$     & 20  \\ \hline
King (CdCC95) Mergers                    & $\alpha = 1.36 \pm 0.08$       & 17  \\  \hline
K Collapses                              & no fit: cluster of data points & 17  \\ \hline 
C Collapses:                             &                                &     \\
n = 0                                    & $\alpha = 2.070 \pm 0.123$     & 10  \\
n = 1                                    & $\alpha = 2.161 \pm 0.087$     & 19  \\ 
n = 2                                    & $\alpha = 2.033 \pm 0.342$     & 10  \\ \hline
A Collapses                              & $\alpha = 1.954 \pm 0.123$     & 22  \\ \hline
AS1 Collapses (all)                      & $\alpha = 2.306 \pm 0.250$     & 9  \\ 
AS1 Collapses (removing AS1-09)          & $\alpha = 2.204 \pm 0.158$     & 8  \\ \hline
AS3 Collapses (all)                      & $\alpha = 2.190 \pm 0.349$     & 9  \\ 
AS3 Collapses (removing AS3-09)          & $\alpha = 1.966 \pm 0.270$     & 8 \\\hline \hline
\multicolumn{3}{c} {Two-component Models:} \\ \hline \hline
Z models ($a_{H} = 10 ~a_{L}$):           &                                &      \\
luminous comp.                           & $\alpha = 1.022 \pm 0.046 $    &  13   \\
dark comp.                               & $\alpha = 1.872 \pm 0.152 $    &  13   \\
both components                          & $\alpha = 1.176 \pm 0.070 $    &  13   \\ \hline 
Zb models ($a_{H} = 3 ~a_{L}$):          &                                &      \\
luminous comp.                           & $\alpha = 1.004 \pm 0.123 $    &  9   \\ \hline
Zc models ($a_{H} = 5 ~a_{L}$):          &                                &      \\
luminous comp.                           & $\alpha = 1.017 \pm 0.105 $    &  9   \\ \hline
\end{tabular}
\end{table}

\subsection{Spin Analysis}

In this section we briefly analyse the how the final spin of the models
depend on the initial condition. We parametrize the spin by the
dimensioness quantity $\lambda$, defined by (c.f. \citealt{pee71}):
\begin{equation}
\lambda = {L |E|^{1/2} \over G M^{5/2}},
\end{equation}
where $L$ is the total angular momentum of the system about its baricenter,
$E$ the total energy of the system, and $M$ the total mass (as already mentioned,
$G=1$).

\begin{figure}
\epsfig{file=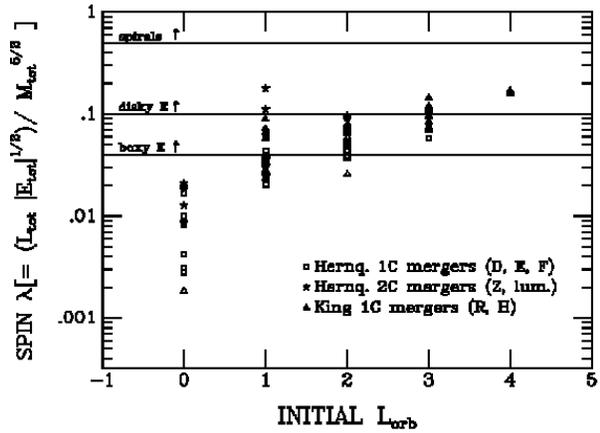,width=8cm}
\caption{Distribution of the spin of the
mergers as a function of the initial orbital angular momentum
of the pre-merger pair. \label{lambda.Lmer}} 
\end{figure}

Fig. \ref{lambda.Lmer} shows how the spin of the
mergers distribute as a function of the initial orbital angular momentum
of the pre-merger pair. First, it can be seen that indeed there is a
transfer of $L_{orb}$ to the final spin of the merger, since higher
$L_{orb}$'s produce systematically higher final spins. Second,
intermediate $L_{orb}$'s ($1 ~<~ L_{orb} ~<~ 3$)
produce objects with spins compatible with boxy ellipticals.
We note, however, that the position of the merger products on the FP
depends very little on $L_{orb}$ (c.f. CDCC95). In other words,
$L_{orb} = 3$ mergers could perfectly be produced from $L_{orb}=0$
mergers, and the final products would have approximately the same positions
on the FP.

%

\begin{figure}
\epsfig{file=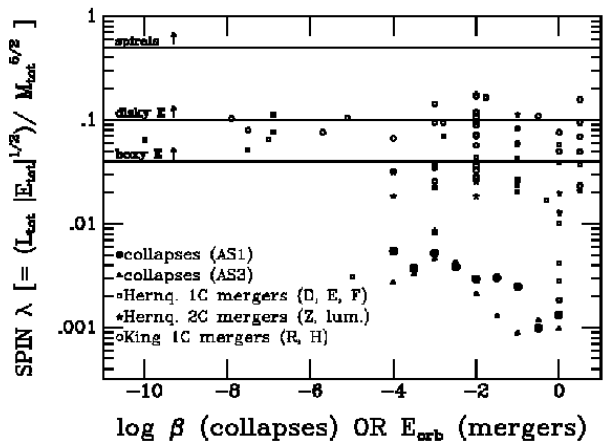,width=8cm}
\caption{Distribution of the spin of collapses and
mergers as a function of the initial collapse factor
($\beta$) and orbital energy ($E_{orb}$), respectively.  \label{lambda}} 
\end{figure}

Fig. \ref{lambda} plot both mergers and collapses as a function of the
initial conditions $E_{orb}$ and $\beta$, respectively. It can be seen
that mergers from a wide range of $E_{orb}$'s are able to produce
objects in the observed range of ellipticals, as opposed to collapses,
which fail in this respect.  It is interesting to notice that
``colder'' collapses reach a higher degree of final spin than the
``hottest'' ones. This seem to imply that the initial rotational
pertubations are amplified in the ``coldest'' collapses. Yet, as we
have seen, these ``colder'' objects still manage to become
approximately homologous (see fits for $AS1-09$ and $AS3-09$ sequences
in Table \ref{tab_ajus_fus}).  Evidently, these results must be
interpreted with caution, since we did not reconfigure the initial
structure of the progenitor in order to include higher initial spins.

\subsection{The Virial Coefficients}

We use another diagnostics for testing homology on the final
simulated objects. A good quantitative measure in this case 
is the direct computation of the kinematical-structural or virial
coefficients ($C_r, C_v$), as described in CdCC95 and DCdCR01:
\begin{equation}
C_r \equiv r_G/r_e
\end{equation}
and
\begin{equation}
C_v \equiv \langle v^2 \rangle / \sigma_0^2;
\end{equation}
where $r_e$ is the effective radius (the radius that defines a sphere
containing half of the total luminosity of the system):  $L(<r_e) =
L_{tot}/2$. $\sigma_0$ is the central projected velocity
dispersion, and $r_G$ is the gravitational radius, defined by
$r_G \equiv GM^2/|W|$, where $W$ is the total potential energy of the
system. $I_e \equiv {L(<r_e)/\pi r_e^2}$ is the mean surface
brightness within $r_e$, in linear units.  Then, $I_e = C_I \left (
{M/2 \over \pi r_e^2} \right )$, with $C_I \equiv \left ( {M \over L}
\right )^{-1}$.  Inserting the equations above into the virial relation
($\langle v^2\rangle = {GM/r_G}$), we find that $r_e = C_{vir}
\sigma_0^2 I_e^{-1}$, where:
\begin{equation}
C_{vir} \equiv {C_r C_v \over 2 \pi G C_I}.
\end{equation}

\begin{figure}
\epsfig{file=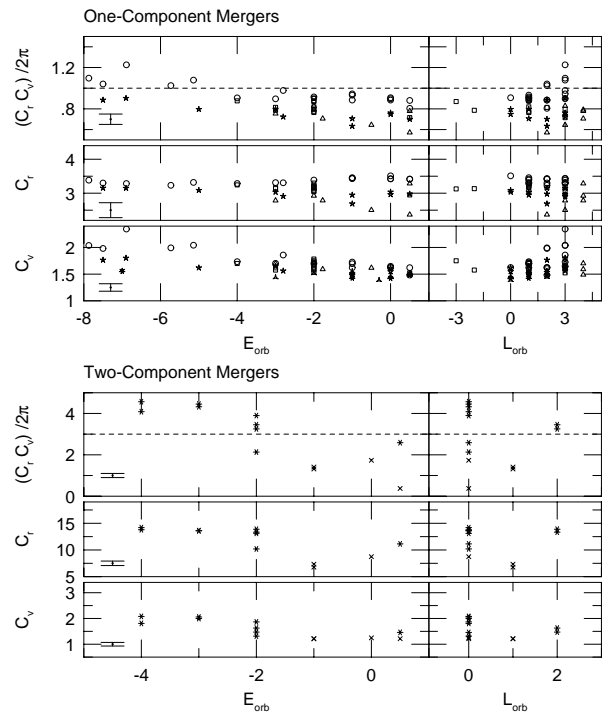,width=8cm}
\caption{Ratio of the virial coefficients
$C_{vir}$, $C_r$ and $C_v$
as a function of the initial conditions.
Symbols are: 5-point stars for the D models; 3-point stars for
the E models; 7-point stars for the E models; 6-point asterisks for the Z
models (1st. gen.); 4-point asterisks the for Z models (2nd. gen.);
open circles for the King models (1st. gen.); open squares
for the King models (2nd. gen.); and open triangles for the
King models (3rd. gen.). A dashed horizontal line is indicated
and represents an arbitrary  homologous family of objects
for comparison.
\label{homol}}
\end{figure}

Since, by construction, $C_I$ (viz. $M/L$) is constant among the
models, the computation of $C_r$ and $C_v$ directly gives the measure
of non-homology among the simulated models.  Note that for
two-component systems, $r_G$ and $\langle v^2 \rangle$ are
calculated from, respectively, the total potential and kinetic energy
of the system. Values of $\sigma_0$ and $r_e$ correspond, however,
only to the visible/barionic matter. As already pointed out,
non-homologous objects are those which the kinematical-structural
coefficients assume different values for each object.  The results are
presented in Fig. \ref{homol}, where we plot the coefficients as a
function of the initial conditions.

First, we find that the structural coefficients, $C_r$, attain
different ranges of values for one and two-component models:  for
one-component mergers, $2.5 {~}^{<}_{\sim}~ C_r {~}^{<}_{\sim} ~3.5$;
whereas for two-component mergers, $8 {~}^{<}_{\sim}~ C_r
{~}^{<}_{\sim} ~15$. This difference is due to the presence of the
massive halo in the two-component models, which pushes the
gravitational radius to larger values, as compared to the one-component
systems. This increase of $r_G$ cannot, however, be compensated by
$r_e$, which depends only on the structure of the luminous core. The
kinematical coefficients, $C_v$, on the other hand, show similar ranges
for both types of mergers. The product $C_r C_v$ (c.f. upper panel of
Fig. \ref{homol}) therefore attain larger values for two-component models than for
one-component ones.

A more relevant aspect of Fig. \ref{homol} is  the fact that the
kinematical/structural coefficients vary in a systematic manner as a
function of the initial orbital energy of the merging models, which is
in agreement with the results found by CdCC95. This behaviour seem to
be an important feature distinguishing mergers from collapses.
Indeed, collapses as a whole are approximately homologous objects,
although some distinctions between ``cold'' and ``hot'' collapses are
found (detailed discussion for collapses can be found in DCdCR01).
There seems to be no correlation with the orbital angular momentum, as
can be seen from an inspection of Fig. \ref{homol}.  On the other hand, it can be
seen that the deviation from homology is more accentuated for
two-component mergers: If we take the total fractional difference
of $y(E_{orb}) \equiv (C_rC_v)/2\pi$, $\delta y \equiv |y(E_{orb~ max})
- y(E_{orb~ min})| / y(E_{orb~ max})$, we find  $\delta y \sim 0.9$ for
two-component mergers whereas $\delta y \sim 0.5$ for one-component
models ($\delta y$ is $\sim 0$ for homologous objects).  This quantity
therefore reproduces the deviation from homology as pictured in the FP
space (c.f. Fig. \ref{fp-lum}), with the advantage that it is possible to trace
the source of non-homology from the corresponding fractional
differences of the $C_r$ and the $C_v$ coefficients separately. For
one-component mergers, $\delta (C_r) \sim 0.14$, $\delta (C_v) \sim
0.25$; for two-component mergers, $\delta (C_r) \sim 0.38$, $\delta
(C_v) \sim 0.42$. Therefore, for one-component mergers, $C_v$
contributes more to the non-homology than $C_r$, a feature that can be
seen clearly from an inspection of Fig.  \ref{homol}.

\subsection{The Ratio of ``Central'' to ``Envelope'' Kinetic Energies}

The results of the previous section demonstrate
that the (central) non-homology effect which characterizes our merger
simulations has a predominant kinematical origin. Now we will analyse the
behavior of the total kinetic energy {\it interior} to a given radius
as compared to the corresponding kinetic energies {\it exterior} to
that radius. In other words, if we call $K_{tot} (<r)$ the ``central''
kinetic energy of the the system and $K_{tot} (>r)$ the kinetic energy
of its ``envelope'', then a measure the ratio of these quantities,$K_x
\equiv K_{tot}(r>xr_h) /K_{tot}(r<xr_h)$, {\it normalized to its
progenitor value}, should reveal, at least in a gross sense, the
effects of the process of relaxation. This process will therefore be
viewed as alterations of the kinetic energies of the more
gravitationally bounded (``central'') particles against the less
bounded ones (``envelope''). Thus if $K_x = 1$ then the end product
model presents the stratification of kinetic energies similar to the
progenitor model.  If however $K_x < 1$ then the ``central'' particles
are ``hotter'' than the ``envelope'', {\it as compared to the
progenitor}.

\begin{figure}
\epsfig{file=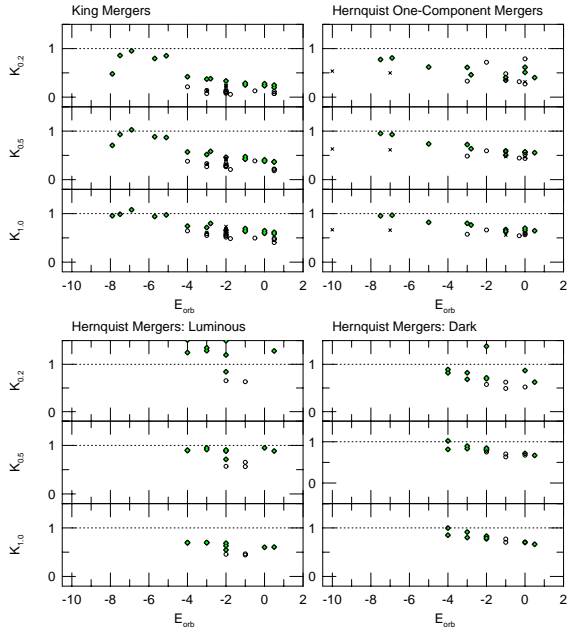,width=8cm}
\caption{Ratio of the kinectic energies exterior and interior to three
distinct radii: $0.2 r_h$, $0.5 r_h$, e $1 r_h$ [notation of the figure:
$K_{x} \equiv K_{tot}(r>xr_h) /K_{tot}(r<xr_h)$, normalized by the
corresponding value of the progenitor (unperturbed model)], as function
of the initial conditions.  The dashed line indicates that the value of
$K_x = 1$. The symbols are: diamonds for the first generation, circles
for the second generation, and crosses for the third generation.
\label{mare.mergers}}
\end{figure}

\begin{figure}
\epsfig{file=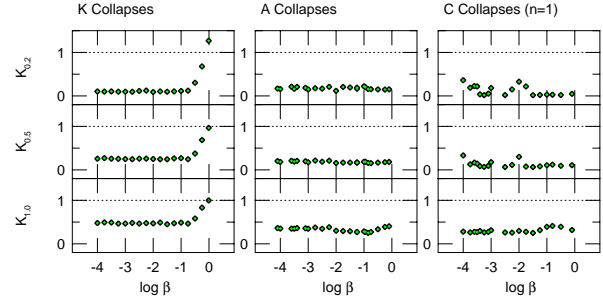,width=8cm}
\caption{The same as previous figure, but for collapses. \label{mare.col}}
\end{figure}

We analysed the kinetic energy ratio, $K_x$, as a function of the
initial conditions (collapse factor or orbital energies, for the
mergers), for three different radii ($x = 0.2$, $0.5$ and $1$). The
results are shown in Fig. \ref{mare.mergers} for the mergers models
and in Fig. \ref{mare.col} for the collapse models.

Most collapse models present $K_x < 1$ for any $x$. 
The ``hottest'' K collapses on the other hand approach $K_x \rightarrow 
1$. In other words, the values of $K_x$ do not change with the 
initial collapse factor ($\beta$), except for the ``hottest'' K collapses.
Moreover the values of $K_x$ are similar among different models, except, 
possibly, the C models which seems more noisy than the others. For larger 
$x$, all the collapse models have $K_x \rightarrow 1$. The general trend is 
that the collapse models are centrally ``hotter'' than the corresponding 
progenitor, \textit{independently of the initial $\beta$} (except for 
$\beta$'s very close to $1$) and the initial model used. 
 
The stratification of kinetic energies in the case of mergers is not
similar to the collapses.  For mergers, it is clear that {\it $K_x$ is
a systematic function of the initial orbital energy of the pairs}. In
other words, for mergers with more negative initial orbital energies
$K_x \rightarrow 1$, showing no difference with their progenitors,
whereas the ones with less negative energies deviate more from the
progenitor, \textit{ and in a systematic way}, towards $K_x < 1$.  The
magnitude of the deviation from $K_x = 1$  also depends on the
merging models: for instance, is greater for the King models,
intermediate for the Hernquist one-component models, and smaller for
the Hernquist two-component models. It also seems to slightly increase
for increasingly merger generations. There are also examples where $K_x
> 1$ (some Hernquist two-component models, with $x=0.2$). In other
words, the behavior of $K_x$ for mergers seems to be more complex than
collapses and shows a clear systematic dependency on the initial
orbital energy of the pairs, in the same sense that the virial
coefficients depend systematically on $E_{orb}$.

In case of mergers, the systematic dependency of $K_x$ on $E_{orb}$
begins to flatten and tend to be erased for sufficiently large values
of $x$. This in fact shows that the merger models tend to a similar
stratification of the ``central'' and ``envelope'' kinetic energies at
sufficiently large radius. In other words, the different $K_x$ values
among the merger models are not only a function of the initial orbital
energy but is a function of $x$ as well, so that the correlation
$K_x \times E_{orb}$ is stronger at the very center of the models and
tend to disappear at sufficiently large radii. This shows therefore
that the effect is intimately related to the central parts of the
system.

Our detailed description of the
ratio of kinetic energies behaviour among models, as given in this section,
seem to reinforce the idea that the non-homolgy in mergers is
a central effect ruled by how the particles gain kinetic energy during
the merger. In other words, the non-homology seem to have a 
dynamical origin which is not present in simple collapses.

In the following, we focus on the analysis of the relaxation history of 
both mergers and colapses, which may help us to find clues for understanding
the dynamical processes that are at the origin of the non-homology of
mergers.

\subsection{Relaxation Histories}

\subsubsection{ Evolution of the Virial Ratio $2K/W$}

\begin{figure}
\epsfig{file=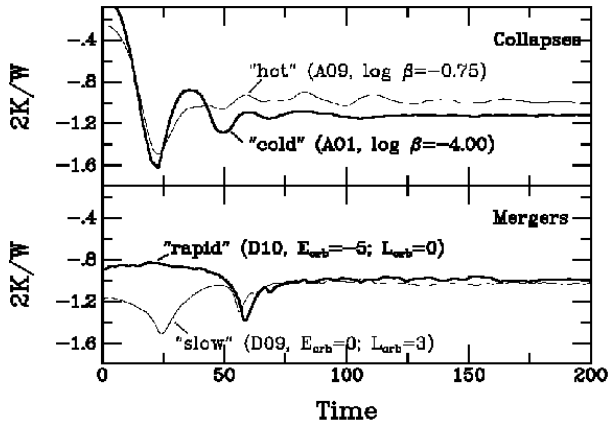,width=8cm}
\caption{The behaviour of the
virial ratio $2K/W$ during the evolution of different and representative
types of models. {\it Upper panel:} collapses. {\it Lower panel:} mergers. 
\label{virial2}} 
\end{figure}

In order to trace a measure of the fluctuations of the gravitational
potential on its way to equilibrium, we compared the behaviour of the
virial ratio $2K/W$ (measured for the whole system, including escapers)
during the evolution of different and representative
types of models, namely: a ``cold'' (A01 model, $\log \beta=-4$) and a
``hot'' (A09 model, $\log \beta=-0.75$) collapse; against a ``rapid'' (D10
model, $E_{orb}=-5$, $L_{orb}=0$) and a "slow" (D9 model, $E_{orb}=0$,
$L_{orb}=3$) one-component merger. Fig. \ref{virial2} shows the behaviour
of these representative models. Notice that some models do not stabilize
around $2K/W = -1$, as would be expected for a virialized model. This is
due to the fact that we are measuring the virial coefficient using
the complete particle data, including particles with positive energies 
which have escaped the system. More ``violent'' relaxations produce more
escapers, and the resulting virial ratio stabilizes around some other value
slightly different than $-1$.

We notice that the most rapidly merging system suffer only one
major fluctuation of $2K/W$, subsequently rapidly reaching equilibrium.
The slower merger shows that $2K/W$ varies in large periods during the
first moments of the evolution (in other words: it does not show a
unique abrupt change in $2K/W$, but rather two or more large periodic
fluctuations before reaching equilibrium). Collapses, regardless of
being ``cold'' or ``hot'' show one large initial fluctuation amplitude
in $2K/W$. Interestingly, subsequent evolution seem to be different:
the ``cold'' collapse still experiences one more relatively significant
fluctuation of $2K/W$ before reaching equilibrium. The ``hot'' collapse,
on the other hand, show a persistent, although of low amplitude,
fluctuation of $2K/W$ still for some time, when the ``colder'' collapse
is comparatively well stabilized.

\subsubsection{The `Kandrup' Effect}

In order to understand the dynamical behaviour presented by the
simulated models, we apply a diagnostic advocated by \citet{kan93}.  The
merging of stellar systems occurs because of a transfer of the orbital
energy to the particles of the stellar systems in question.  The
mechanisms through which this occurs are the tidal interactions, which
increase the internal energy of the systems at the expense of
their orbital energy. The question here concerns the relation of this
mechanism with the central non-homology of the simulated mergers.

During the evolution of the system, the energy of the particles is not,
in general, conserved, even in a ``coarse grained'' sense (viz. through
the distribution $N(E)$; for a discussion on the importance of this
distribution for stellar systems, see \citealt{bin82}).  Kandrup et al.
studied the distribution of the energy of the particles in systems
resulting from collisions (without the formation of a final single
object) and merging of two galaxies.  These authors found that {\it
there is} a ``coarse-grained'' sense in which the {\it ordering} of the
mean energy of given collections of particles is unaltered, even though
$N(E)$ may vary substantially.  In this section, we revisit the
question raised by Kandrup et al. and try to connect this fact to the
behaviour of the simulated systems in context of the FP.  Notice that
their conclusions were based on only two simulations of collisions,
with only one merger, and two collapses. Here we use a much larger set
of simulations and initial conditions, and a larger number of particles
as compared to the models used by Kandrup et al.  We will not consider
time evolutions of mean energies, as Kandrup et al. did, but only the
initial and final values of the mean energy of the given collections of
particles. We discuss further their diagnostic below.

\begin{figure}
\epsfig{file=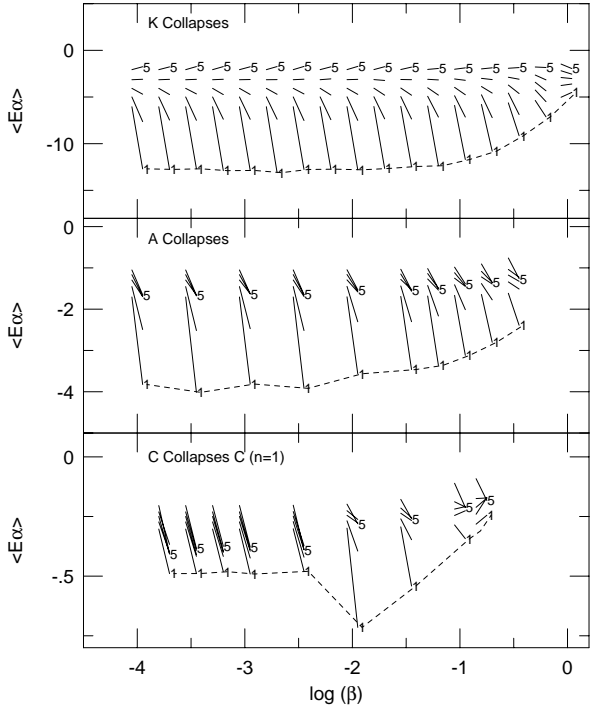,width=8cm}
\caption{Energy means of $5$ equal number collections of particles,
ordered accordingly to their initial mean binding energies.
The initial and final mean binding energy values are connected by a line
segment (left extreme: initial mean value; right extreme:
final value), centered in the initial condition of each model.
``Bin numbers'' are indicated in the figure: bin number
$1$ refers to the most bounded particles; and successively 
throughout bin number $5$, which refers to the most weakly
bound particles. The values of the final mean binding energy
of the initially most bounded bin (1) are connected with
a dashed line, illustrating how it changes as a function of the
initial condition. The models shown are collapses. \label{fa4.5.col}}
\end{figure}

\begin{figure}
\epsfig{file=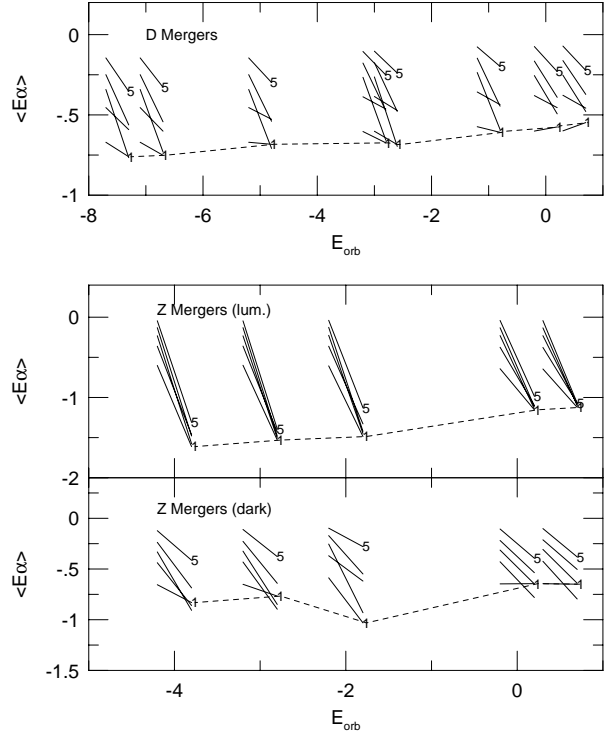,width=8cm}
\caption{The same as previous figure, but for first generation mergers. In the
case of mergers with equal $E_{orb}$'s, we have included in the figure
only the model with lower $L_{orb}$.
\label{fa4.5.mer}}
\end{figure}

The method may be considered a ``lagrangian'' approach to the analysis
of how the energy of the particles change because of the relaxation
process.  The particles of the \textit{initial} models have been sorted
accordingly to their binding energies and the models were partitioned
into 5 bins of equal number of particles (a finer partitioning with 10
bins produced essentially the same results). For each of these
bins, the mean energy was calculated and the bins ranked with the
first one initially containing the most bound
particles (most negative mean energy) whereas the fifth, the less
bounded ones (less negative mean energy). The mean energy of these
collections of particles were then recalculated at the end of the run's
and compared with their initial values. We have limited our analysis
for the first generation of mergers.  In the case of mergers with equal
$E_{orb}$'s, we have included only the model with lower $L_{orb}$. The
results of these comparison are displayed in Figs. \ref{fa4.5.col} and
\ref{fa4.5.mer}.

We found that, except for some of the C cases, collapses preserve the ordering of the mean
energies per bin entirely. These results confirm the findings
of Kandrup et al. Moreover the mean energies per bin that changed
more in this case were the ones corresponding to the most bounded bins
(1, 2, 3, etc). The central potential becomes deeper after the
collapse, and the particles initially more bounded to the system tend
to loose energy, becoming even more tied. This effect is also a
function of the collapse factor $\beta$, as can be seen by the dashed
line in the figure, which connects the most bounded bin (1),
illustrating how this bin changes as a function initial condition. In
the case of A collapses, the mean energies representing the three less
bounded bins converge to similar final values, whereas the two most
strongly bounded bins reach even more negative mean values. This
behaviour indicate that collapses tend to produce core-halo structures.  In
the case of C collapses, the mean energies change considerably and
chaotically. Recall that these models contain a Hubble flux which may
favor the grow of perturbations embedded in these models, adding some
complexity in the evolution of the mean energy of these distinct
collections of particles.

In the case of mergers (one and two-component models), the preservation
of the ordering of the mean energies per bin is not as good as in the
case of collapses.  For the D models (Hernquist one-component models),
the initially most bounded particles will remain as the most bounded
ones after the relaxation process. However, it can be seen that the bins
number 3 actually cross bins number 2 and reach the values
corresponding to bins number 1.  In the case of two-component models,
the luminous matter tend to reach very negative values of the mean
energies, almost converging to similar values for all bins. The general
behaviour of the luminous component resemble the behaviour of the most
boundly tied particles of the collapse models.  The main reason for
this may be the fact that, after the initial interaction, the luminous
component finds its equilibrium state within the deeper potential well
of the dark halo.  This might occur through a partial collapse of the
luminous matter inside the dark halo.  In the case of the dark matter
component, there is also some violation of the ordering of the mean
energies per bin.  In fact, the initially most boundly tied bins (number
1) crosses upwards and gain energy in some cases.  The halo seems to be
the only system that actually shows clearly this behaviour.

The D models and the luminous component also present the same effect as
seen in collapses:  the particles initially more bounded to the system
tend to loose energy as a function of the initial orbital energy, as
can be seen by the dashed line in the figure, which connects the most
bounded bins (1).  For halos, this behaviour is not as clear.

Note that the mean biding energy of the most bounded bins 
remains at an almost constant value for
collapses (see dashed lines in Fig.  \ref{fa4.5.col}), rising steeply
for the ``hottest'' collapses. The C models present more fluctuations
in this behaviour. For mergers, on the other hand, these changes
proceed more smoothly and
systematically with the initial orbital energy  (dashed lines in Fig.
\ref{fa4.5.mer}).  This means that the initially most bounded
collection of particles remain at the average the most bounded
particles after the relaxation, but at a more negative mean energy than
the slower (less negative $E_{orb}$) mergers.

It is clear that if the ordering of the mean energies of
particles, partitioned at a coarse-grained level, is strictly
conserved, as in collapses, then the most bounded particles (in average
closer to the baricenter) continue be the most bounded particles after
virialization.  However, as pointed out previously, this ordering
conservation {\it does not} occur for mergers. 
In other words, some complex behaviour seem
to take place during the merger involving the more central or bounded
particles, an effect which {\it does not} occur at all in collapses
(except for some small shuffling between the {\it less} bounded bins
for the ``hottest'' C models). In summary, we entirely confirm
the results of Kandrup et al. for collapses, but in the case
of mergers, some violation of the ordering conservation of the
energy bins is present.

\section{Discussion}

We have simulated a hierarchical non-dissipative merger scheme similar
to that of CdCC95, however using different models for representing the
progenitors of the $\rm 1^{st}$ generation mergers. In contrast with
CdCC95, which considered King density profiles, we used models endowed
with cuspy profiles, such as the Hernquist density profile. Also models
with a dark halo second component were used in this study. A comparison
with collapse simulations (previously analysed in DCdCR01 and here
extended to include collapses with initial spins) is presented.

We found that the one-component Hernquist mergers give results similar
to those found by CdCC95 for the one-component King models, namely both
were able to build-up small scattering FP-like correlations with slopes
consistent to those found for the near infrared FP of nearby galaxies.
The two-component models also reproduce a FP-like correlation, but with
a significantly steeper slope which is in agreement with that found for
galaxies at high redshift (\citealt{pah98a}). Pahre finds that the
slope of the near-infrared FP decreases with increasing redshift (see
his Figure 7.2).  Another important piece of evidence of the evolution
of the FP with redshift comes from the work of \cite{kel97}. The
authors find that the structure of the galaxies in the analysed sample
has not changed significantly since  $z = 0.58$, based on the fact that
the observed scatter is rather low: $\pm 0.067$ in $\log r_e$.
Besides, they find a dependence between $M/L_V$ and redshift, which
reinforces the idea of a stellar population effect in the evolution of
the FP.

In nature, dissipational effects must have played a role in producing
the FP relations and their scatter, but it would be unwise to
completely disregard the role of stellar dynamics in shaping
ellipticals as well.  Our simulations are only of a dynamical
character, with the $M/L$ ratio fixed by construction. Systematic
non-homology in the evolved models can produce FP-like `tilts'
compatible with those found in nature.  In particular, we show the
importance of the gravitational potential of the halo for changing the
`tilt' of the FP: the magnitude of the change can be seen directly from
a comparison between one-component and two-component Hernquist merger
models (c.f. Fig. 1). In other words, this simple result clearly shows
how the FP slope may be dynamically changed just by the addition of a
halo. Therefore, our results suggest that the structure evolution of
the halo could also have a collateral importance in changing and
shaping  the FP `tilt', along with population evolution effects (e.g.
\citealt{kel97}).  We speculate on the possibility that halos may suffer
evolution from $z \sim 0.5$ to the present.  The evolution could be in
the form of tidal stripping, which would decrease the mass of the halo,
or by the presence of supermassive central black holes, which could
alter the matter distribution of the halo, forming a core (c.f.
\citealt{hen01}).  If dark matter presents some level of
self-interaction (c.f. \citealt{spe00}), then it may drive evolution
towards a core in less than a Hubble time (c.f.  \citealt{yos00}). If
some of these processes have operated in halos during the past few
Gyrs, in the sense of altering their gravitational potentials at the
centers of galaxies, the characteristic scaling properties of the
luminous/barionic component might have changed as well. Whether any of
these possibilities are in factible is at present an open issue.

A qualitative analysis of the behaviour of the mean energy of
collections of particles (as advocated by Kandrup et al. 1993) lead us
to consider the possibility that `mesoscopic' constraints could have
some connection to the central non-homology. The conservation of the
ordering of the mean energy of collections of particles implies that
the process of `mixing' in the one-particle energy space is quite
inefficient as compared with `mixing' in configuration and/or
velocity space (see discussion in Kandrup et al.). This seem to be true
for collapses, but not entirely for the central parts of mergers.

The most intense tidal perturbations (shocks) seem to be found for the
most rapidly merging systems (more negative orbital energies).  In this
case, the particles probably withdraw the energy from the relative
orbit of the merging pairs at one major fluctuation.  Secondary
fluctuations on the gravitational potential evolve afterwards rapidly
and reach equilibrium in a short timescale as well. The stratification
of the kinetic energies resemble that of the progenitor in this case.
On the other hand, if the orbital energy is less negative (slow
mergers), there is some of time for the particles to withdraw energy
from the orbit of the pair, and this process involves periodically
large fluctuations on the potential that evolve slowly, taking a larger
amount of time to stabilize. This process may be important in ``heating
up'' the central parts of the models approaching $E_{orb} \rightarrow
0$.  This should be important in defining the non-homology in mergers
because the stratification of the kinetic energies are indeed different
to that of the progenitor.
 
On the other hand, in the case of collapses, the dynamics seem to
operate in a different manner than in mergers.  Collapses starts off
from a spherically symmetric condition that mergers do not share.
Collapses also produce not only fast by very high amplitude
gravitational potential fluctuations that dump rapidly.  This process
should be very efficient in heating up the central parts of the models
in configuration and/or velocity space, but not efficient enough to
have the particles `forget' their initial energies in a collective
(`mesoscopic') manner.  As already pointed out by Kandrup et al., this
behaviour is at odds with Lynden-Bell's theory of `violent
relaxation', where `mixing' in energy space is not expected to be
inefficient for any given collection of particles.  At the same time,
we have found that collapses seem to `prefer' forming homologous
systems, whereas mergers do not. Some connection between `mesoscopic'
constraints and non-homology seem to be apparent, but this is an open
question.

We did not atempt at this time to rigorously try to connect the
behaviour of the violation of the ordering of the energy bins (Figs.
\ref{fa4.5.mer} and \ref{fa4.5.col}) with the behaviour of the
`central' to `envelope' kinetic energies (Figs. \ref{mare.mergers}
and \ref{mare.col}).  Although interesting, in order to fully
understand this effect, we would need to probe the problem of
relaxation in a much deeper and/or formal manner, which is not the
objetive of our paper at the present time.  Figs. \ref{fa4.5.mer} and
\ref{fa4.5.col} illustrate a diagnostic on the behaviour of the change
of energy of the system due to relaxation in a `mesoscopic' scale.
Figs.  \ref{mare.mergers} and \ref{mare.col} show a different
dignostic, where the change of energy of the central and external parts
of the system are compared to their {\it progentor models}, not the to
their initial condition (as in Figs.  \ref{fa4.5.mer} and
\ref{fa4.5.col}), and hence refer to a more `macroscopic' feature of
the relaxation process. On the other hand, the main point of Figs.
\ref{fa4.5.mer} and \ref{fa4.5.col} is {\it not} to show how the energy
of collections of particles change due to relaxation (although it also
certainly shows that) but in {\it what degree} the ordering of their
mean energies is violated.  This type of analysis was first envisaged
by Kandrup in 1993 and is still not well understood.  In our oppinion, it is
not at all clear how one could find any immediate connection between
both sets of figures. We know that the non-homology comes primarily as
a systematic function of $E_{orb}$.  Both sets of figures show
systematic behaviour of 2 different types of diagnostics as a function
of $E_{orb}$. Collapses show almost no dependecy of these same
diagnostics with $\beta$ (except for the ``hottest'' collapses).
Therefore, it seems that relaxation through merging embodies some
mechanism which is effective in differentiating the final models,
producing non-homology, whereas this mechanism is absent or highly
precluded in collapses (again, except for the ``hottest'' ones, which
are just a small perturbation form equilibrium of the progenitor
model).  In fact, what is lacking in order to make any progress in this
direction is a systematic understanding of the nature of the
gravitational relaxation mechanism, where several conceptual issues are
still unsolved (c.f. \citealt{pad90}).

In any case, our results seems to strenghten the idea that
dissipationless merging could produce significant non-homology in the
final objects and therefore FP-like relations in the same sense and
with comparable values of the FP `tilt' as those observed in
ellipticals.  We have shown that, from purely dynamical grounds,
mergers can produce FP-like relations while {\it simple} collapses
cannot (two-component collapses were not investigated here and will be
a subject for future work). The evolution of gradients in the
gravitational field of the merging galaxies seem to dictate the final
non-homology of the end products. Further investigations are necessary
in order to stablish, quantify and rigorously explain these complex
effects, as preliminary discussed in this paper.

\section*{Acknowledgements}
We thank J. Dubinski and R. Carlberg for usefull discussions during
this project.  We also thank the anonymous referee for his/her
constructive suggestions. C.C.D. acknowledges fellowships from FAPESP
under grants 96/03052-4 and 01/08310-1.  A.L.B.R.  acknowledges
fellowships from FAPESP under grant 97/13277-6. This work was partially
supported by CNPq and PRONEX-246.

\bsp

\label{lastpage}

\end{document}